\documentclass[amsmath,amssymb,aps,prd,showpacs,onecolumn]{revtex4}
\usepackage{graphics}
\usepackage{epsfig}
\usepackage{hyperref}

\begin{document}
\title{Triangular Ring Resonator: Direct measurement of the  parity-odd parameters of the photon sector of SME}
\author{Qasem Exirifard }
\affiliation{Physics Department, K.~N.~Toosi University, Tehran, Iran \\
Physics School, Institute for Research in Fundamental Sciences (IPM),
Tehran, Iran\\
exir@theory.ipm.ac.ir}
\begin{abstract}
We introduce the  the Triangular Ring  (TR) resonator. We show that the difference between the clockwise and anti-clockwise resonant frequencies of a vacuum TR resonator is sensitive to the birefringence parity-odd parameters of the photon's sector of the minimal Standard Model Extension (mSME): the Standard Model plus all the perturbative  parameters encoding the  break the Lorentz symmetry. We report that utilizing the current  technology  allows for  direct measurement of these parameters with a sensitivity of the parity even ones and improves the best current resonator bounds by couple of orders of magnitudes.   \\
We  note that  designing an optical table that rotates perpendicular to  the gravitational equipotential surface (geoid) allows  for direct measurement of   the constancy of the light speed at the vicinity of  the earth in all directions in particular perpendicular to the geoid. If this table  could achieve the precision of the ordinary tables, then it would improve the GPS bounds on the constancy of the light speed perpendicular to geoid by about eight orders of magnitude. 
\end{abstract}
\maketitle


\section{Introduction}	
 Currently nature, \textit{a capite ad calcem}, is perceived in the doctrine of the three paradigms: Special Relativity, General Relativity, and Quantum Mechanics.  This doctrine suffers from the problem of quantum gravity.  Not withstanding the success of this doctrine, perhaps we must quantitatively  question  how precise nature nurtures each of these paradigms.   Prior to  an appropriate reply, test models for the paradigm of the question  must be constructed. In this work we  consider  a test model for the special relativity.

The most recent test model for the special relativity is the Standard Model Extension (SME) \cite{SME0,SME1,SME2}. This model adds all the perturbative parameters breaking the Lorentz symmetry to the Lagrangian of the standard model before asking how much which experiment or observation constrains them.  This model has stemmed   recent research aimed to detect or study the Lorentz Invariance Violation (LIV) terms for electromagnetic (photon sector of the SME) in various fields, including the  classical solutions of SME electrodynamics \cite{Kobakhidze:2007iz,Casana:2008sd0,Casana:2008sd1,Bailey:2004na},  cosmological birefringence constraints \cite{Kostelecky:2001mb0,Kostelecky:2001mb1,Exirifard}, radiation spectrum of the electromagnetic waves and CMB data \cite{Kahniashvili:2008va,Lue:1998mq,Feng:2006dp,Wu:2008qb,Kostelecky:2007zz,Kamionkowski:2008fp,Balaji:2003sw},  black body radiation in finite temperature \cite{Casana:2009xf,Casana:2009dq,Casana:2008ry,Fonseca:2008xa},  LIV terms in higher dimensional scenarios \cite{Gomes:2009ch,Baukh:2007xy},  synchrotron radiation \cite{Bocquet:2010ke,Gurzadyan:2010bt}, Cherenkov radiation \cite{Altschul:2006zz,Kaufhold:2007qd,Klinkhamer:2007ak0,Klinkhamer:2007ak1}
 and modern cavity resonators or interferometry experiments \cite{Eisele:2009zz,Muller:2007zz,Saathoff:2003zz,NaturePhysics,Tobar:2009gw,Hohensee:2010an}.  Studies have been conducted to search for the LIV terms in the neutrino sector \cite{SMENeutrino0,SMENeutrino1,SMENeutrino2,SMENeutrino3,SMENeutrino4,SMENeutrino5,SMENeutrino6},  meson sector \cite{SMEMesons0,SMEMesons1,SMEMesons2,SMEMesons3,SMEMesons4,SMEMesons5,SMEMesons6,SMEMesons7, SMEMesons8, SMEMesons9,SMEMesons10}, electron sector \cite{SMEelectron0,SMEelectron1,SMEelectron2,SMEelectron3,SMEelectron4,SMEelectron5,SMEelectron6,SMEelectron7,SMEelectron8,SMEelectron9,SMEelectron10,SMEelectron11,SMEelectron12}, proton sector \cite{SMEproton0,SMEproton1,SMEproton2,SMEproton3,SMEproton4}, neutron sector \cite{SMEneutron0,SMEneutron1,SMEneutron2,SMEneutron3,SMEneutron4,SMEneutron5,SMEneutron6,SMEneutron7,SMEneutron8,SMEneutron9} and also the gluon sector \cite{Carone:2006tx}. The reader is advised to look at the data table of all the SME parameters \cite{Kostelecky:2008ts}.   So far no compelling non-zero value for the LIV  parameters has been reported \cite{ Kostelecky:2008ts}  ``though SME-based models provide simple explanations for certain unconfirmed experimental results including anomalous neutrino oscillations \cite{Noscillations} and anomalous meson oscillations \cite{Moscillations}" \cite{Finsler}.

In this work we consider the CPT even part of the minimal SME,  the photon sector of which reads:
\begin{eqnarray}
{\cal L} &=& -\frac{1}{4} F_{\mu\nu} F^{\mu\nu}\,-\,\frac{1}{4} (k_{F})_{\mu\nu\lambda\eta} F^{k\lambda} F^{\mu\nu}\,,\nonumber
\end{eqnarray}
where $k_F$'s components are small real numbers.   We then report that a Triangular Fabry-Perot resonator (which we will introduce) is capable of measuring the  parity odd components of the $k_F$ tensor. We show that  utilizing a vacuum TR resonator on the current rotating tables will improve the  direct bounds on five out of the eight parity-odd parameters (at the scales of 10 eV ) by about four orders of magnitude. The remaining three parameters can be measured by a TR resonator partially filled with a magnetic material. 

The paper is organized as follows:  we first provide a short glimpse on the mSME electrodynamics in the vacuum and in presence of matter. We then introduce the Triangular Fabry-Perot resonator and present our motivation to introduce this resonator. We, next,  show  that a non-vacuum TR resonator measures the parity-odd non-birefringent parameters of the mSME. In its following section we derive with what precision a vacuum TR resonator can measure, this time, the birefringent parity-odd parameters. Before we summarize we talk about the possibility of new tests for QED in GR.

\section{Electrodynamics of mSME}
In order to address the SME electrodynamics, it is more common to express $k_F$ tensor in term of  new $3\times 3$ matrices defined by 
\begin{eqnarray}
\left( \kappa_{DE}\right)  ^{jk} & \equiv&-2(k_F)^{0j0k}\,,\\
\left(\kappa_{HB}\right)^{jk}&\equiv&\frac{1}{2}\epsilon^{jpq}\epsilon^{klm}(k_F)^{pqlm}\,,\\
\left(  \kappa_{DB}\right)  ^{jk} & \equiv&-\left(  \kappa_{HE}\right)^{kj}=\epsilon^{kpq}(k_F)^{0jpq}\,.
\end{eqnarray}
The equations of motion for SME electrodynamics  can be expressed in the form of modified source free Maxwell equations:
\begin{eqnarray}
\nabla. D & =& 0 \,,\\
\nabla . B & =& 0\,, \\
\nabla \times E + \partial_t B & =& 0\,, \\
\nabla \times H - \partial_t D & =& 0\,,
\end{eqnarray}
where the modified definition of $D$ and $H$ read
\begin{equation}
\left(\begin{array}{c} D \\ H \end{array}\right)
\,=\, \left(
\begin{array}{cc}
\epsilon_0 \kappa_{DE} & \sqrt{\frac{\epsilon_0}{\mu_0}} \kappa_{DB} \\
\sqrt{\frac{\epsilon_0}{\mu_0}} \kappa_{HE} & \mu_0^{-1} \kappa_{HB}
\end{array}
\right)
\left(\begin{array}{c} E \\ B \end{array}\right)\,.
\end{equation}
Note that $\kappa_{DE}$, $\kappa_{DB}$, $\kappa_{HE}$ and $\kappa_{HB}$ matrices are not the eigenfunctions of the parity and time reversal operator. The eigenfunctions of these operators read
\begin{eqnarray}
(\tilde{\kappa}_{e+})^{jk} & =& \frac{1}{2} (\kappa_{DE} + \kappa_{HB})^{jk}\,,\\
(\tilde{\kappa}_{e-})^{jk} & =& \frac{1}{2} (\kappa_{DE}- \kappa_{HB}) -\frac{1}{3} \delta^{jk} (\kappa_{DE})^{ll} \,,\\
(\tilde{\kappa}_{o+})^{jk} & =& \frac{1}{2} (\kappa_{DB}+ \kappa_{HE})^{jk}\,,\\
(\tilde{\kappa}_{o-})^{jk} & =& \frac{1}{2} (\kappa_{DB}- \kappa_{HE})^{jk}\,,~(\tilde{\kappa})=\frac{1}{3} (\kappa_{DE})^{ll}\,,
\end{eqnarray} 
where the subscript of $\pm$ represents the eigenvalue under time reversal, and the subscripts of $o$ and $e$ means the oddness or evenness   under the parity operator. Note that $\tilde{\kappa}$ is a scalar, all matrices are traceless, and all  matrices are symmetric but $\tilde{\kappa}_{o+}$ which is antisymmetric.

Later in this work, we shall need mSME in the presence of the matters.  For mSME electrodynamics in the presence of matter we follow the notation of \cite{Tobar:2004vi}, and in the line of \cite{Kostelecky:2001mb1} where 
\begin{equation}
\left(\begin{array}{c} D \\ H \end{array}\right)
\,=\, \left(
\begin{array}{cc}
\epsilon_0 (\tilde{\epsilon}_r + \kappa_{DE}) & \sqrt{\frac{\epsilon_0}{\mu_0}} \kappa_{DB} \\
\sqrt{\frac{\epsilon_0}{\mu_0}} \kappa_{HE} & \mu_0^{-1} (\tilde{\mu}_r^{-1}+ \kappa_{HB})
\end{array}
\right)
\left(\begin{array}{c} E \\ B \end{array}\right)\,,
\end{equation}
In vacuum $\tilde{\epsilon}$ and $\tilde{\mu}$ are identity matrices. In this paper we consider isotropic materials so 
\begin{eqnarray}
\tilde{\epsilon} & =& \epsilon_r {\bf 1}\,,\\
\tilde{\mu} & =& \mu_r {\bf 1}\,.
\end{eqnarray} 
A general $k_F$ leads to birefringence in vacuum. If we assume that the universe is homogeneous then the infrared, optical, and ultraviolet spectropolarimetry of various cosmological sources at distances  $0.04-2.08 Gpc$ \cite{data10,data11,data12,data13,data14,data15,data16,data17,data18,data19} bound the components of $\tilde{\kappa}_{e+}$ and $\tilde{\kappa}_{o-}$ to less than $2 \times 10^{-32}$ at $90\%$ confidence level \cite{SME0,SME1}. This combined with the modern Michelson-Morley  experiments \cite{Eisele:2009zz,Muller:2007zz} then demands $\tilde{\kappa}_{e-}$ and $\tilde{\kappa}_{o+}$ to be less than $10^{-16}$ at $90\%$ confidence level \cite{Exirifard}. Perhaps it is interesting to obtain the limits on $k_F$ without assuming that the universe is homogeneous. In a non-homogeneous universe we can not deduce direct (model-independent) information on the local variables  from the cosmological observations. In we abandon the assumption of a homogenous universe, we should consider local experiments to directly constrain the LIV parameters.

The local constraints can be provided by accelerators. For example not having observed the Compton-edge photons at the ESRF's GRAAL facility \cite{Bocquet:2010ke,Gurzadyan:2010bt}  leads to the bound of $10^{-13}$  on the mSME coefficients. There exist proposals to measure the parity-odd LIV terms at the level of $10^{-16}$ by the electrostatics or magnetostatics systems \cite{Kobakhidze:2007iz,Casana:2008sd0,Casana:2008sd1}.  Interferometry system  lead to local constraints as well. The most recent  Michelson-Morley-type experiment \cite{Eisele:2009zz}, improving previous bounds \cite{Muller:2007zz},  reports the bound of $10^{-17}$  on the parity even. In the SME the Parity-odd LIV terms mix with the parity-even ones under boost. Due to the orbit of the earth around the Sun, therefore,  a limit on the parity-odd terms can be deduced for the parity odd coefficients from the results  of the Michelson-Morley experiments. Doing so, the sensitivity for the parity-odd terms, is four orders of magnitude less than that of the parity-even ones. This means that  \cite{Eisele:2009zz} implies the limit of  $10^{-13}$  on the parity  odd coefficients. It is interesting to design and introduce a cavity resonator that is sensitive to the parity-odd terms. The possibility for designing these kinds of resonators is discussed in  Ref. \cite{Mewes:2006ad}. This paper aims to introduce a simple  resonator which is directly sensitive to (all or some of) the parity-odd LIV terms of the photon sector of the SME. 
 
 \section{Triangular Fabry-Perot Resonator}
\begin{figure}[tbp]
\centering
\psfig{file=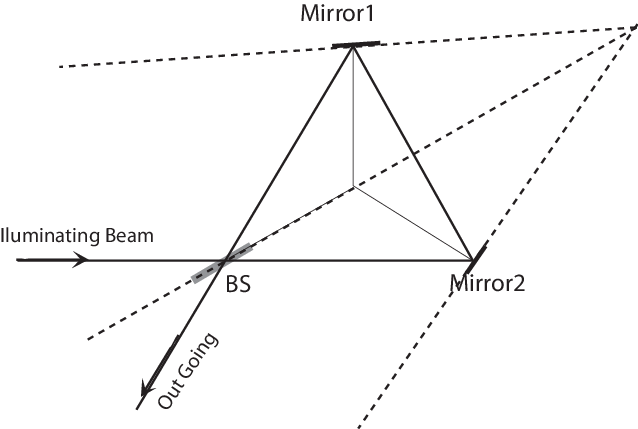,width=7cm}
\vspace*{8pt}
\caption{Triangular Interferometry: The fringe pattern between light rays moving clockwise and ant--clockwise is measured, reproduced from ref. \cite{Triangle}}
\label{fig:1}
\end{figure}

The Michelson-Morley  experiment \cite{Michelson0,Michelson1} measures the two-way light speed: the average of the light speed in opposite directions. In the early  test models of the special relativity such as the Robertson-Mansouri-Sexl Model \cite{Robertson,Mansouri}, the intrinsic deviation from special relativity was encoded only in the two-way light speed. SME in contrary to the Robertson-Mansouri-Sexl Model has detectable  parity-odd parameters: parameters that affect the one-way light speed. The parity-odd parameters of the SME can/may be directly detected using  new interferometry/resonator systems. In such an interferometer the light path should be closed.  In order to avoid taking the average of the light speed in the opposite directions, no segment of this path should be parallel to any other of its segments. A triangular interferometry system wherein light moves on the perimeter of a triangle possesses these properties: fig. [\ref{fig:1}]. The triangular interferometer was first introduced by Trimmer \textit{et. al}  \cite{Trimmer}. It appears that it was rediscovered later in optics \cite{Triangle}: as a  system which is robust to the fluctuations caused by the environment.  Nowadays people in laboratories  often use Mach-Zehnder  interferometer in order to systematically suppress the fluctuations. We note that a vacuum Mach-Zehnder interferometer/resonator  measures only the two-way light speed. As we will show a vacuum triangular  resonator can measure some of the parity-odd parameters.

In 1973, Trimmer \textit{et. al} used the triangular interferometry and reported the vanishing of  the parity-odd terms (albeit within his test model for special relativity not the SME) with the precision of $10^{-10}$ \cite{Trimmer}. The Trimmer experiment, however, has not yet been repeated or improved. Here we suggest to repeat and improve this experiment. In so doing, we first change  the triangular interferometer to the triangular resonator.

\begin{figure}[tbp]
\centering
\psfig{file=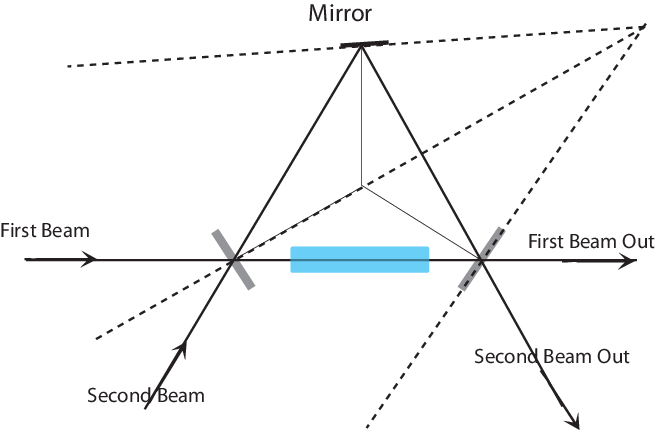,width=7cm}
\vspace*{8pt}
\caption{Triangular Resonator:   A  triangle resonator can produce multiple resonant frequencies.}
\label{fig:2}
\end{figure}

A setup for the triangular resonator is depicted in Fig. \ref{fig:2}. This setup uses a perfect mirror and two other mirrors which partially allow light rays pass through them. Similar to the Trimmer setup \cite{Trimmer}, we have placed a piece of glass (a magnetic material, $\mu>1$ ) in one edge of the resonator in order to make it sensitive to $Y_{1m}(\theta,\phi)$ parity-odd LIV terms. The resonator is transparent to frequencies in which  a standing light wave is produced on the perimeter of the triangle. In order to calculate the resonant frequencies, we just need to find the waves which allow the formation of the standing light wave inside the resonator.  

Triangular resonator is a simple generalization of the Fabry-Perot resonator. Fig. \ref{fig:2} depicts how to choose the resonant frequencies of lights moving clockwise  and anti-clockwise using a single triangular resonator. In the next section we prove that measuring the beat frequency between these two resonant frequencies  as the optical table rotates, will measure the parity-odd parameters of the photon sector of SME. 

\section{mSME Electrodynamics and TR resonator}
Ref.  \cite{Eisele:2009zz} reports that the parity-even LIV terms in the mSME are zero with the precision of parts in $10^{17}$.  So we set 
\begin{eqnarray}
\label{1a}
(\tilde{\kappa}_{e+})^{jk} & =&0,\\
\label{2a}
(\tilde{\kappa}_{e-})^{jk} & =& 0,
\end{eqnarray} 
at the precision we are working in this paper. Ref. \cite{Altschul:2009xh} reports that   $|\kappa| < 5 \times 10^{-15}$. $|\kappa|$ affects the anisotropy of light at level of $|\kappa| \frac{v}{c}$ where $v$ stands for the velocity of the laboratory with respect to an inertial frame. The effect of the isotropic parameter, thus, is about $10^{-19}$. We want to address the isotropy of the light speed with the precision of $10^{-17}$. So we set
\begin{equation}
\label{3a}
\tilde{\kappa} \,=\, 0\,.
\end{equation} 
This leaves us the eight parity odd parameters: $(\tilde{\kappa}_{o+})^{jk}$, $(\tilde{\kappa}_{o-})^{jk}$. The five components of $(\tilde{\kappa}_{o-})^{jk}$ contribute to the first order birefringence. So they are called the birefringent terms. The three components of the $(\tilde{\kappa}_{o-})^{jk}$ does not cause birefringece at the leading order so they are called non-birefringent terms. In the following sub-sections we first show how TR resonator can measure the non-birefringent terms, and in its subsequent subsection we show how a TR resonator can measure the birefringent terms. 

\subsection{Non-Birefringent parity-odd parameters}
\begin{figure}[tbp]
\centering
\psfig{file=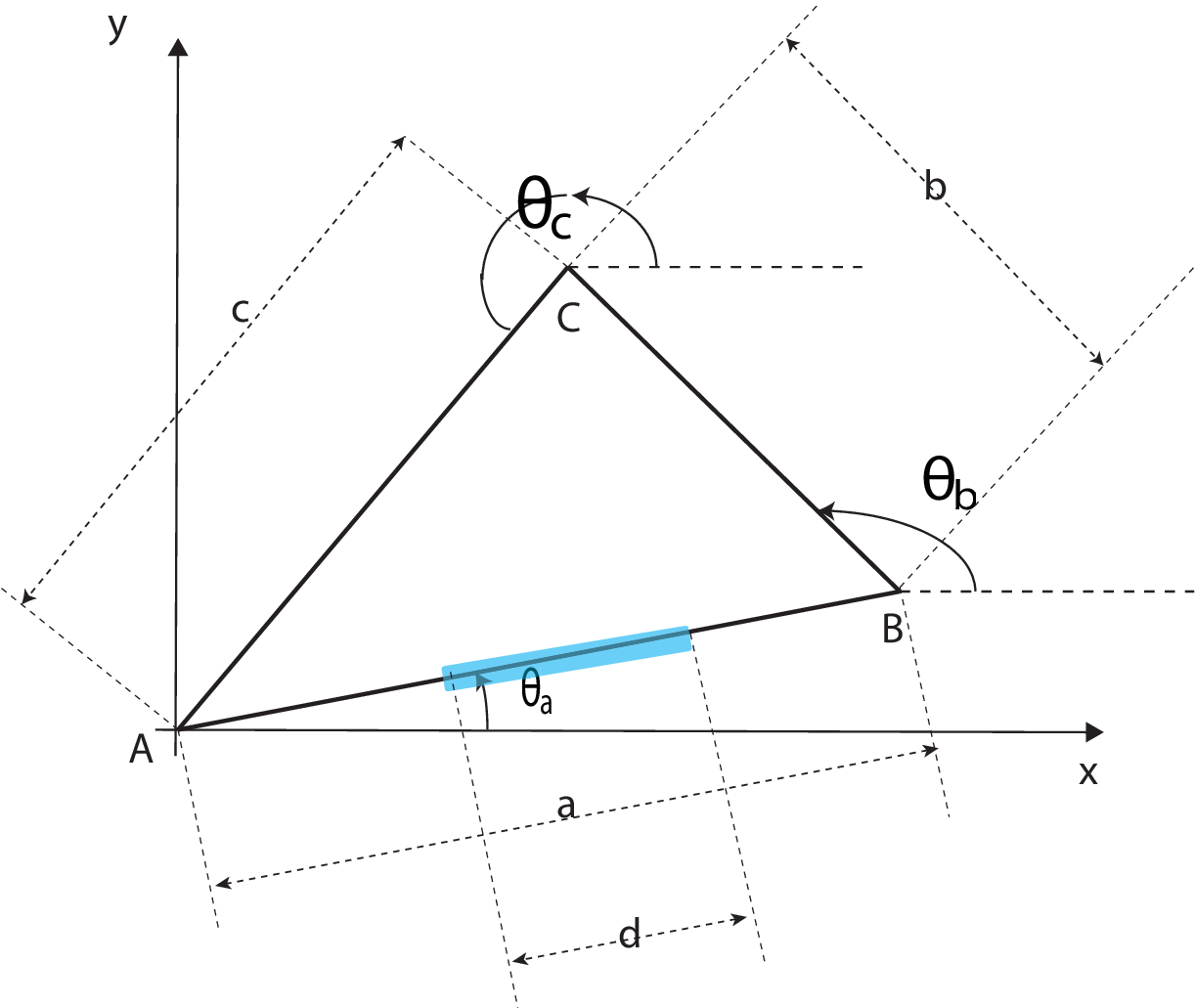,width=7cm}
\vspace*{8pt}
\caption{A triangular resonator and how  we label it. }
\label{fig:4}
\end{figure}

Table VIII of ref. \cite{Kostelecky:2009zp} has listed the matrix elements relating the cartesian to spherical coordinates of the CPT-even and parity-odd of the photon sector of the minimal SME. Examining these tables illustrates that the three components of  $(\tilde{\kappa}_{o-})^{jk}$  are mapped to the spherical component of angular momentum ($l$) one. The contribution of $(\tilde{\kappa}_{o-})^{jk}$ to the light speed reads:
\begin{equation}
\frac{{\bf c}}{c(\theta,\phi)}  \,=\, 1  - p(\theta,\phi)\,,
\end{equation}
where ${\bf c}$ is the light speed in the special relativity, which now can be written by
\begin{eqnarray}
\label{tml}
 p(\theta,\phi) & =& \sum_{m}  p_{m}Y_{1m}(\theta,\phi)\,,
\end{eqnarray}
where $p_m$ are the components of the  $(\tilde{\kappa}_{o-})^{jk}$ in the spherical coordinates of a universal rest frame. 
Now consider  an arbitrary triangular resonator.  Choose the coordinates such that the resonator is placed on the $xy$ surface. Denote the angle between the positive direction of the $x$ axis and $\vec{AB}$ edge of the triangle by $\theta_a$, fig. \ref{fig:4}:  there exists a transparent isotropic material in one arm of the triangle, for now please just consider a vacuum resonator. We assume that we can rotate the resonator in the $xy$-plane in order to test the anisotropy of the light speed in the $xy$-plane. We search for anisotropy in the $xy$-plane. As the earth rotates, the $xy$-plane probes all the space. So, providing a formulation for the light anisotropy in the $xy$-plane suffices to test the anisotropy of the light speed in all directions.
Consider an inertial frame attached to the laboratory.  A general anisotropy for the light  speed in vacuum in the $xy$-plane reads
\begin{equation}\label{ctheta}
\frac{{\bf c}}{c(\theta)} \,=\, 1 + p(\theta)\,,
\end{equation} 
wherein $\theta$ is the angle between the light's direction and the  positive $x$ direction. Furthermore projecting \eqref{tml} on the $xy$ plane yields 
\begin{equation}\label{ctheta1}
 p(\theta)\,=\, q \sin (\theta + \theta_0)\,,
\end{equation} 
where $q$ is a function of $p_m$ and the configuration of the triangle with respect to the universal rest frame. 

Now present the resonant frequency of the anti-clockwise rotating beam (the first beam of fig. \ref{fig:2} without a material inside it) by $\nu^+$.  This resonant frequency holds
\begin{equation}\label{nu+}
\frac{d}{c(\theta_a)} + \frac{b}{c(\theta_b)} + \frac{c}{c(\theta_c)} \,=\, \frac{n^{+}}{\nu^+}\,,
\end{equation}
where $n^{+}$ labels the $\nu^+$ modes.  Due to \eqref{ctheta1} we have
\begin{eqnarray}\label{NoGoVacuum}
a\, p(\theta_a) + b\, p(\theta_b) + c\, p(\theta_c)\,=\,0\,.
\end{eqnarray}
which shows that  $(\tilde{\kappa}_{o-})^{jk}$ does not contribute  to the resonant frequencies of the 
the vacuum TR resonator.  No change in the perimeter  of the vacuum resonator makes it sensitive to   $(\tilde{\kappa}_{o-})^{jk}$ \cite{Trimmer}\footnote{\cite{Trimmer} has proved this statement for an interferometer. Extending this proof to the kind of the resonators we are considering here is trivial.}. 

So let us now consider a non-vacuum resonator: Fig.  [\ref{fig:2}]. To derive the resonant frequency of a non-vacuum resonator we should derive the light speed for mSME in the presence of matter. Let an isotropic matter be considered.  Then ref. \cite{Tobar:2004vi} has presented the light speed for mSME in the presence of matter.    Eq. (16) of the ref. \cite{Tobar:2004vi} beside its appendix A,  shows that at the leading order in the parameters, ignoring the corrections due to boost we have 
\begin{eqnarray}\label{kup}
\frac{k^{\uparrow}_{xy}}{\sqrt{\mu_r \epsilon_r} k_0} 
& = &  
1 +\sqrt{\frac{\mu_r}{\epsilon_r}} (\kappa_{DB})^{xy}\,,\\
\label{kdown}
\frac{k^{\downarrow}_{xy}}{\sqrt{\mu_r \epsilon_r} k_0} & = &  1 - \sqrt{\frac{\mu_r}{\epsilon_r}} (\kappa_{DB})^{xy} \,,
\end{eqnarray}
the superscript refers to the direction of the propagation along the z axis ($\uparrow$ positive $\downarrow$ negative), the subscripts labels the direction of the $E$ and $H$ field and $k_0$ is the constant of the propagation in the vacuum in the absence of LIV coefficients.  Note that we have simplified eq. (16) of the ref. \cite{Tobar:2004vi} for the purpose of this section. In the sense that \eqref{1a}, \eqref{2a} and \eqref{3a} have been implemented in \eqref{kup} and \eqref{kdown}.

We notice that  \eqref{kup} and \eqref{kdown}  lead to expressing  the parity odd coefficients  in the material in term of the parity coefficients in the vacuum. To have this illustrated recall:
\begin{eqnarray}
k^{\uparrow}_{xy} c_{xy}^\uparrow &=& \omega\,,\\
k^{\downarrow}_{xy} c_{xy}^\downarrow&=& \omega\,,
\end{eqnarray}
where $c_{xy}^\uparrow$ and $c_{xy}^\downarrow$ represent the light speed in the opposite directions along the z axis. Then it follows that 
\begin{equation}\label{vnv}
\frac{{\bf c}_m} {c_{xy}^\uparrow} - \frac{{\bf c}_m} {c_{xy}^\downarrow} \,=\, 2 \sqrt{\frac{\mu_r}{\epsilon_r}} (\kappa_{DB})^{xy}\,,
\end{equation}
which dictates that the parity odd expansion in \eqref{tml} in presence of matter holds
\begin{equation}
p_m^{\text{mat}} \, =\, p_m\, \sqrt{\frac{\mu_r}{\epsilon_r}}   \,,
\end{equation}
where $p_m$ is the coefficient in the vacuum while $p_m^{\text{mat}}$ stands for the coefficient in a given  isotropic material.

Now consider  an arbitrary triangular resonator with an isotropic matter in one of its arm.  Choose the coordinates such that the resonator is placed on the $xy$ surface: fig. \ref{fig:4}. We assume that we can rotate the resonator in the $xy$-plane in order to test the anisotropy of the light speed in the $xy$-plane. We again search for anisotropy in the $xy$-plane. 

The SME light speed inside the isotropic material in $xy$-plane, $c_m(\theta)$, is directional dependence and it reads
\begin{equation}\label{cmSME}
\frac{{\bf c}_m}{c_m(\theta)} \,=\, 1 + \sqrt{\frac{\mu_r}{\epsilon_r}} p(\theta)\,,
\end{equation} 
wherein $\theta$ is the angle between the light's direction and the  positive $x$ direction and $p(\theta)$ is given in  \eqref{ctheta1} and ${\bf c}_m$ represents the light speed in the isotropic material in the ordinary electrodynamics:
 \begin{equation}
 {\bf c}_m = \frac{1}{\sqrt{\mu_r \epsilon_r}} \textbf{c}\,,
 \end{equation}
 Now present the resonant frequency of the anti-clockwise rotating beam (first beam of fig. \ref{fig:2}) by $\nu^+$.  This resonant frequency holds
\begin{equation}\label{nu+duplicate}
\frac{a-d}{c(\theta_a)} +\frac{d}{c_m(\theta_a)} + \frac{b}{c(\theta_b)} + \frac{c}{c(\theta_c)} \,=\, \frac{n^{+}}{\nu^+}\,,
\end{equation}
where $n^{+}$ labels the $\nu^+$ modes. 
For sake of simplicity define:
\begin{eqnarray}
\hat{a} & =& a + (\sqrt{\mu_r \epsilon_r}-1) d \,.
\end{eqnarray}
Inserting \eqref{ctheta} into \eqref{nu+duplicate} yields
\begin{eqnarray}\label{nu++}
\frac{{\bf c}n^{+}}{\nu^+} & =&  \hat{a}+b+c +\\
& &+(a+ (\mu_r-1) d)\, p(\theta_a) + b\, p(\theta_b) + c\, p(\theta_c)\,.\nonumber
\end{eqnarray}
Using  \eqref{NoGoVacuum} simplifies \eqref{nu++} to
\begin{eqnarray}\label{nu+++}
\frac{{\bf c}\,n^{+}}{\nu^+} & =&  \hat{a}+b+c + (\mu_r-1) d\, p(\theta_a) \,.\nonumber
\end{eqnarray}
We denote the resonant frequency of the clockwise rotating beam by $\nu^-$ (the second beam of fig. \ref{fig:2}). $\nu^-$ satisfies 
\begin{equation}\label{nu-}
\frac{\hat{a}}{c(\pi+\theta_a)} + \frac{b}{c(\pi+\theta_b)} + \frac{c}{c(\pi+\theta_c)} \,=\, \frac{n^{-}}{\nu^-}\,,
\end{equation}
where $n^{-}$ labels the modes for $\nu^-$. Inserting \eqref{ctheta} into \eqref{nu-} yields
\begin{eqnarray}\label{nu--}
\frac{{\bf c}\,n^{-}}{\nu^-} & =&  \hat{a}+b+c - (\mu_r-1) d\, p(\theta_a)\,.\nonumber 
\end{eqnarray}
Let the resonant frequencies be expressed in term of the average frequency, $\bar{v}$ and their beat frequency: 
\begin{eqnarray}
\label{nu+a}
\frac{1}{\nu^{+}} & =& \frac{1}{\bar{\nu}} (1 +\frac{1}{2} (\epsilon(\theta_a)+\epsilon_0)) \,,\\
\label{nu-a}
\frac{1}{\nu^{-}} & =& \frac{1}{\bar{\nu}} (1 - \frac{1}{2}(\epsilon(\theta_a)+\epsilon_0)) \,,
\end{eqnarray}
where the beat frequency is defined by
\begin{equation}
\Delta \nu=\bar{\nu}(\epsilon({\theta_a})+\epsilon_0)\,.
\end{equation}
Note that we have
\begin{equation}
\epsilon(\theta_a)+\epsilon_0 << 1\,.
\end{equation}
Also notice that  $\epsilon_0$ is the beat frequency predicted by the  special relativity and $\epsilon(\theta_a)$ represents the deviation from the special relativity's prediction.  Also up to the leading order we can use (and define $\bar{n}$)
\begin{equation}
n^{+} \approx n^{-} = \bar{n}\,.
\end{equation}
Using \eqref{nu+++}, \eqref{nu--}, \eqref{nu+a} and \eqref{nu-a} leads to 
\begin{eqnarray}\label{21}
-\frac{{\bf c}\epsilon(\theta_a)}{\bar{\nu}}  &=&
\frac{2(\mu_r -1)}{\bar{n}} d \,p(\theta_a) \,.
\end{eqnarray} 
Every quantity in the right hand side of  \eqref{21} is observable: $\bar{\nu}$ is the average of the resonant frequencies, $\nu\epsilon(\theta_a)$ is the configuration dependent part of the beat frequency, that is:
\begin{equation}
\bar{\nu}\epsilon(\theta) \,=\, \Delta\nu(\theta) - \Delta\nu(0)\,,
\end{equation} 
$\bar{n}$ is the average of labeling of  the considered modes of the resonator. In summary we get
\begin{equation}\label{22}
\frac{\Delta\nu(\theta) - \Delta\nu(0)}{\bar{\nu}} \,=\, \frac{2 (\mu_r -1) d}{\hat{a}+ b + c}\,p(\theta),
\end{equation} 
Eq. \eqref{22}  makes us enable to measure the  anisotropy of the one-way light speed.  
    
\subsubsection{Precision of the vacuum and matter-filled Triangular resonator}
The setup of ref. \cite{Eisele:2009zz}  and \cite{Muller:2007zz} has reached the precision of $10^{-17}$ for the parity even parameters. We do assume the beat frequency between the clockwise and anticlockwise resonant frequencies of the  vacuum triangular resonator can reach to the precision of ref. \cite{Eisele:2009zz}  and \cite{Muller:2007zz}, that is 
\begin{equation}
\left.\frac{\Delta \nu}{\nu}\right|_{\text{vac.}} < 10^{-17}.
\end{equation}
This assumption is valid since the beat frequency of the clockwise and anti-clockwise resonant frequencies of the   vacuum triangular resonator is insensitive to the environmental noises.  

Compered to the Fabry-Perot resonator, a rotating ring resonator is sensitive to the Sagnac effect \cite{Sagnac1,Sagnac2}. A small deviation from a constant angular velocity  produces a signal in the beat frequency of the clockwise and anti-clockwise resonant frequency of the ring resonator. In the search for the  SME parameters on a rotating table, thus, the fluctuations in the Sagnac effect due to fluctuations of the angular velocity of the rotating table should be  taken into account. Perhaps one way is to design the rotating table such that it rotates with a constant angular velocity with the desired precision. The other way is to cancel the Sagnac signals by data processing. The later is possible since  the Sagnac effect is proportional to the area of the resonator while SME signals are proportional to the perimeter of the ring. Therefore placing two ring resonators of different areas  on  the same rotating table,  and then measuring the beat frequencies of the clockwise and anti-clockwise resonant frequencies of each ring resonator, allows for systematically removing  the Sagnac effect signals by data processing. We suggest to remove the fluctuation of the angular velocity by data processing. Since there exists  a simple way to remove the fluctuations of the Sagnac effect,  the vacuum Triangular resonator indeed can reach at least the precision of  ref. \cite{Eisele:2009zz}  and \cite{Muller:2007zz}.

When we use a material inside the resonator the loss in the material we use in the triangular resonator suppresses the precision of the vacuum resonator 
\begin{equation}
\left.\frac{\Delta \nu}{\nu}\right|_{\text{r}} <  10^{-17}\, {\cal{T}}(d).
\end{equation}
where ${\cal{T}}(d)$ is the transmission coefficient of the material we use in the resonator.   
 Eq. \eqref{22} then indicates that we can measure $p(\theta)$ with the precision of 
 \begin{equation}\label{23}
 p(\theta) < \frac{\hat{a}+ b + c}{2 d (\mu_r -1)} {\cal{T}}(d) \times 10^{-17}\,.
 \end{equation}
To simplify this relation, consider the triangular resonator for which
\begin{equation}
a \,=\, b\,=\, c \,=\, d \,=\, \frac{L}{3}\,,
\end{equation}
where $L$ is the perimeter of the triangle, then \eqref{23} converts to 
\begin{equation}
 p(\theta) < \frac{\sqrt{\mu_r \epsilon_r}+ 2}{2  (\mu_r -1)} {\cal{T}}(\frac{L}{3}) \times 10^{-17}\,.
\end{equation}
So if we can find an almost lossless material for which 
\begin{equation}
\frac{\sqrt{\mu_r \epsilon_r}+ 2}{2  (\mu_r -1)} {\cal{T}}(\frac{L}{3}) < 1 \,,
\end{equation}
the suppression of the precision of the resonator due to loss inside the material is ignorable. The problem is that the author is not aware of any lossless magnetic material in the optical range. Usually $\mu_r$ is larger than one near the resonant frequency of the material where absorption is large too. So it does not sound easy to propose or suggest  a magnetic material  transparent for the frequency used in the setup of \cite{Eisele:2009zz} or \cite{Muller:2007zz}.

There is also  a challenging problem associated with the partially filled  triangular resonator. The experiment requires that only part of the optical path be filled with a material. `There will be reflection (from the clockwise beam to the counterclockwise and vice versa) at the boundaries of the crystal that we have not considered.' We should employ a technique to suppress this reflection. Ref. \cite{Baynes:2011nw}, done and reported after this paper,  has presented a sharp way to suppress this reflection. It uses a ring resonator with a prism at the Brewster's angle. Ref. \cite{Baynes:2011nw} also reports the  precision of $10^{-13}$ for non-rotating ring/triangular resonator. This high precision is inherited from the triangular or Mach-Zhender interferometer systems \cite{Triangle}.  Note that ref. \cite{Baynes:2011nw} has not performed the experiment on a rotating table. We suggest the experiment to be done over a rotating table, albeit  the fluctuation of the Sagnac effect must be taken into account in order to achieve the  precision of \cite{Eisele:2009zz}.  ``In this suggestion following the literature we have assumed that the lengths of the resonator arms remain unchanged under rotations. The apparatus could in principle also be affected by Lorentz-symmetry violations  but it appears likely that they should be anyway very small''
. Notice  that  the triangular resonator still possesses some merits.  Next section shows that the vacuum triangular resonator is sensitive to the birefringent parity odd parameters. We urge the experimentalist to directly measure these parameters using a vacuum triangular resonator. 

\subsection{Birefringent parity-odd parameters}  
Table VIII of ref. \cite{Kostelecky:2009zp}  illustrates that the five components of  $(\tilde{\kappa}_{o+})^{jk}$  are mapped to the spherical component of angular momentum ($l$) two. This implies that the contribution of $(\tilde{\kappa}_{o+})^{jk}$ to the light speed of a given polarization reads:
\begin{equation}
\frac{{\bf c}}{c(\theta,\phi)}  \,=\, 1  - \sum_{m=-2}^{2} \tilde{p}_m Y_{2m}(\theta,\phi)\,.
\end{equation}
The contribution of $Y_{2m}(\theta,\phi)$  to the optical path for a closed path is not vanishing. In the previous subsection we were forced to use the matter-filled TR resonator because the net contribution of $Y_{1m}(\theta,\phi)$ for over a closed path was zero. This is not the case for  $Y_{2m}(\theta,\phi)$. A vacuum TR resonator  measures $\tilde{p}_m $.    
If we use unpolarized beam inside a vacuum TR resonator, then the SME model leads to multiple resonant frequencies for it. The frequency shifts in the vacuum TR resonator due to SME corrections, however,  will be at the order of 
\begin{eqnarray}
|\frac{\Delta \nu}{\nu}|  \propto  \tilde{p}_m  \,.
\end{eqnarray} 
If we measure the frequency of a vacuum TR resonator we then can deduce constraint on $\tilde{p}_m$.

The setup of ref. \cite{Eisele:2009zz}  and \cite{Muller:2007zz} has reached the precision of $10^{-17}$ for the resonant frequency. As argued before, perhaps this precision can be achieved for a vacuum TR. The only difference between the vacuum TR  resonator and ordinary FP resonator is that the TR resonator uses three mirrors instead of two. So we assume that all the state of art technology utilized in the FP resonator to reach the precision of  $10^{-17}$ can be utilized in the TR to reach almost the same precision.  Let it be emphasized again that the beat frequency of the clockwise and anti-clockwise resonant frequencies of the TR compared to that of two FP resonators is insensitive to the environmental noises. The TR resonator does not need the very complex noise reduction system needed to compare the resonant frequencies of two independent FP resonators. 
This means that  TR   will lead to the direct measurement of the birefringent parity-odd parameters with a sensitivity better than parts in $10^{17}$. This improves the direct interferometry constraints on these coefficients reported in \cite{Tobar:2009gw} by  six orders of magnitude, and  the bound of ref. \cite{Eisele:2009zz}  and \cite{Muller:2007zz} on the parity-odd parameters by about  four orders of magnitude. The direct experiment is measuring the parity-odd birefringent parameters without assuming anything on the homogeneity of the universe.  Its outcome, therefore, is complementary  to the cosmological bounds on the parity-odd birefringence parameters reported in \cite{Kostelecky:2001mb0,Kostelecky:2001mb1} wherein a homogeneous universe assumed.

\section{New test for QED in  General Relativity}
The CPT-even part of the mSME for the electromagnetic sector reads
\begin{equation}\label{MGEh}
{\cal L} \,=\, -\frac{1}{4} F_{\mu\nu} F^{\mu\nu}\,-\frac{1}{4} (k_{F})_{\mu\nu\lambda\eta} F^{\lambda\eta} F^{\mu\nu}\,,
\end{equation}
where $k_{F}$ has 19 algebraically independent components. The standard approach assumes  that $k_F$ is constant in the whole of the universe. In a curved space-time geometry  $k_F$ may depend on the space-time geometry \cite{Bailey:2006fd}. It can be a functional of the  Riemann tensor and its covariant derivatives:
\begin{equation}\label{MGEh1}
 (k_{F})_{\mu\nu\lambda\eta}=k_{F}[l_{s}^2 R_{ijkl}, g_{ij}, l_{_{SME}} \nabla_i ]_{\mu\nu\lambda\eta}\,,
\end{equation}
where  $l_{_{SME}}$  plays the role of the SME scale. A consistent fundamental theory for the spontaneous break of the Lorentz  symmetry in a curved space-time geometry would identify $k_F$.  In the absence of such a theory, we can just estimate the constraint on $l_s$. In order to provide this constraint we note that around the Earth it holds
\begin{equation}
|R_{\mu\nu\lambda\eta}| \equiv (R_{\mu\nu\lambda\eta} R^{\mu\nu\lambda\eta})^{\frac{1}{2}} \propto \frac{G M_{\text{Earth}}}{c^2 r_\text{Earth}^3}\propto 10^{-23} \frac{1}{m^2} \,.
\end{equation}
Gravity so far has illustrated two lengths: the planck length and the length of the cosmological constant: 
\begin{eqnarray}
l_{p} & \propto& 10^{-35} m \,, \\
l_{\Lambda} &=&  \frac{1}{\sqrt{|\Lambda|}} \,\propto\,10^{26} m \,,
\end{eqnarray}
where $\Lambda$ is the measured value for the cosmological constant:  the cosmological constant determined by +SN Ia observation reads \cite{Lambda}:
\begin{equation}
|\Lambda| \,=\, 1.206^{+0.064}_{-0.073} \times 10^{-52}\, \frac{1}{m^2}\,.
\end{equation}
If what causes the cosmological constant  also causes the break of the Lorentz symmetry then $l_{_\text{SME}}\propto l_{\Lambda}$ and around the Earth 
\begin{eqnarray}
l_{_\text{SME}}^2 |R_{\mu\nu\lambda\eta}|\propto 10^{29} \,,
\end{eqnarray}
which is very large and will be ruled out immediately. If quantum gravity induces LV than $l_{_\text{SME}}\propto l_p$ and around the Earth we have
\begin{eqnarray}\label{c1}
l_{_\text{SME}}^2 |R_{\mu\nu\lambda\eta}|\propto 10^{-93} \,,
\end{eqnarray}
which is very small. QED has another scale: the Compton wavelength of electron. If this scale plays a role than  ($l_{_\text{SME}}\propto10^{-12} m$) and
\begin{eqnarray}\label{c2}
l_{_\text{SME}}^2 |R_{\mu\nu\lambda\eta}|\propto 10^{-47} \,,
\end{eqnarray}  
which is still very small.  \eqref{c1} and \eqref{c2} imply that we do not  have any theoretical motivation to search for $l_{_{SME}}$. We, however, can provide better and better empirical limits on the possible scale of SME by performing experiments.  All the  implemented experiments  in laboratory have had the table of interferometry system rotated  parallel to the earth surface. So they have not and are not measuring the possible dependency of $k_F$ perpendicular to the gravitational equipotential. It is interesting to design the table of interferometry system such that  the table rotates perpendicular to the earth surface.  Once this table is designed and constructed, a combination of the triangular resonator and Fabry-Perot resonator will constrain the  light speed isotropy near the surface of the earth, in the directions of parallel to the surface and perpendicular to the surface, with the precision of  about $3 \frac{nm}{s}$. This  precision is not sufficient to address quantum gravity but it provides a better limit on the scales where break of the Lorentz symmetry may be observed.  In particular it improves the limit of ref. \cite{GPS} on the parity-even parameters by about eight orders of magnitudes and implies that $l_{_{SME}}<1 km$.

\section{Conclusions}
We have introduced a new resonator: the Triangular Ring resonator.  We have shown that the difference between the clockwise and anti-clockwise resonant frequencies of a vacuum TR resonator measures the birefringence parity-odd parameters of the photon's sector of the minimal Standard Model Extension (mSME).  Utilizing the current  technology  leads to the direct measurement of these parameters with a sensitivity better than parts in $10^{17}$, which is four orders of magnitude improvement.  We also have reported that the partially filled triangular resonator measures the non-birefringence parity-odd parameters of the QED sector of the mSME.

We also have realized that after designing new rotating tables, tables that rotate perpendicular to the equipotential surface, a TR resonator and ordinary Fabry-Perot resonators shall provide new constraints on the form of the non-minimal coupling between curvature of the space-time geometry and QED. They will report whether or not the light speed is the same in all directions near the Earth. The current direct resonator experiments report that the light speed is the same tangent to the equipotential surface (geoid). The  new table shall experimentally measure and would report the constancy of the light speed in all directions with about a precision of $3 nm/s$.

\section{Notes added}
One of the proposed experiments was carried out and the constraints were improved by orders of magnitudes \cite{Michimura:2013kca, Michimura:2013via}.

\section*{Acknowledgments} 
This work was supported by the school of physics, Institute for Research in Fundamental Sciences. I also would like to thank the HECAP section of the Abdus Salam International Centre for Theoretical Physics for the nice hospitality.


\begin{thebibliography}{000}

\bibitem{SME0}
D.~Colladay and V.~A.~Kostelecky,
  Phys.\ Rev.\ D {\bf 55} (1977) 6760
  [hep-ph/9703464].
  
\bibitem{SME1}
  D.~Colladay and V.~A.~Kostelecky,
  Phys.\ Rev.\ D {\bf 58} (1998) 116002
  [hep-ph/9809521].

\bibitem{SME2}
 V.~A.~Kostelecky,
  Phys.\ Rev.\ D {\bf 69} (2004) 105009
  [hep-th/0312310].


\bibitem{Kobakhidze:2007iz}
  A.~Kobakhidze and B.~H.~J.~McKellar,
  Phys.\ Rev.\ D {\bf 76} (2007) 093004
  [arXiv:0707.0343 [hep-ph]].
  
\bibitem{Casana:2008sd0}
 R.~Casana, M.~M.~Ferreira, Jr and C.~E.~H.~Santos,
  Phys.\ Rev.\ D {\bf 78} (2008) 105014
  [arXiv:0810.2817 [hep-th]].
  
\bibitem{Casana:2008sd1}
  R.~Casana,  M.~M.~Ferreira, A.~R.~Gomes and P.~R.~D.~Pinheiro,
Europ. Phys. Jour. C {\bf62} (2009) 573 [arXiv:0812.1813 [hep-th]].
  
\bibitem{Bailey:2004na}
Q.~G.~Bailey and V.~A.~Kostelecky,
  Phys.\ Rev.\ D {\bf 70} (2004) 076006
  [hep-ph/0407252].

\bibitem{Kostelecky:2001mb0}
 V.~A.~Kostelecky and M.~Mewes,
  Phys.\ Rev.\ Lett.\  {\bf 87} (2001) 251304
  [hep-ph/0111026].
  
\bibitem{Kostelecky:2001mb1}  
   V.~A.~Kostelecky and M.~Mewes,
  Phys.\ Rev.\ D {\bf 66} (2002) 056005
  [hep-ph/0205211].
  
 \bibitem{Exirifard}
 Q.~Exirifard,
  Phys.\ Lett.\ B {\bf 699} (2011) 1
  [arXiv:1010.2054 [gr-qc]].


\bibitem{Kahniashvili:2008va}
T.~Kahniashvili, R.~Durrer and Y.~Maravin,
  Phys.\ Rev.\ D {\bf 78} (2008) 123009
  [arXiv:0807.2593 [astro-ph]].
  
 
\bibitem{Lue:1998mq}
A.~Lue, L.~-M.~Wang and M.~Kamionkowski,
  Phys.\ Rev.\ Lett.\  {\bf 83} (1999) 1506
  [astro-ph/9812088].


\bibitem{Feng:2006dp}
 B.~Feng, M.~Li, J.~-Q.~Xia, X.~Chen and X.~Zhang,
  Phys.\ Rev.\ Lett.\  {\bf 96} (2006) 221302
  [astro-ph/0601095].
    
\bibitem{Wu:2008qb}
   E.~Y.~S.~Wu {\it et al.}  [QUaD Collaboration],
  Phys.\ Rev.\ Lett.\  {\bf 102} (2009) 161302
  [arXiv:0811.0618 [astro-ph]].
  
\bibitem{Kostelecky:2007zz}
  V.~A.~Kostelecky and M.~Mewes,
  Phys.\ Rev.\ Lett.\  {\bf 99} (2007) 011601
  [astro-ph/0702379 [ASTRO-PH]].
  
\bibitem{Kamionkowski:2008fp}
  M.~Kamionkowski,
  Phys.\ Rev.\ Lett.\  {\bf 102} (2009) 111302
  [arXiv:0810.1286 [astro-ph]].

\bibitem{Balaji:2003sw}
K.~R.~S.~Balaji, R.~H.~Brandenberger and D.~A.~Easson,
  JCAP {\bf 0312} (2003) 008
  [hep-ph/0310368].



\bibitem{Casana:2009xf}
  R.~Casana, M.~M.~Ferreira, Jr and M.~R.~O.~Silva,
  Phys.\ Rev.\ D {\bf 81} (2010) 105015
  [arXiv:0910.3709 [hep-th]].

\bibitem{Casana:2009dq}
   R.~Casana, M.~M.~Ferreira, Jr, J.~S.~Rodrigues and M.~R.~O.~Silva,
  Phys.\ Rev.\ D {\bf 80} (2009) 085026
  [arXiv:0907.1924 [hep-th]].
  
 
\bibitem{Casana:2008ry}
  R.~Casana, M.~M.~Ferreira, Jr. and J.~S.~Rodrigues,
  Phys.\ Rev.\ D {\bf 78} (2008) 125013
  [arXiv:0810.0306 [hep-th]].
  
  
\bibitem{Fonseca:2008xa}
J.~M.~Fonseca, A.~H.~Gomes and W.~A.~Moura-Melo,
  Phys.\ Lett.\ B {\bf 671} (2009) 280
  [arXiv:0809.0704 [hep-th]].


\bibitem{Gomes:2009ch}
  M.~Gomes, J.~R.~Nascimento, A.~Y.~.Petrov and A.~J.~da Silva,
  Phys.\ Rev.\ D {\bf 81} (2010) 045018
  [arXiv:0911.3548 [hep-th]].
  
\bibitem{Baukh:2007xy}
  V.~Baukh, A.~Zhuk and T.~Kahniashvili,
  Phys.\ Rev.\ D {\bf 76} (2007) 027502
  [arXiv:0704.0314 [hep-ph]].
  
\bibitem{Bocquet:2010ke}
 J.~-P.~Bocquet, D.~Moricciani, V.~Bellini, M.~Beretta, L.~Casano, A.~D'Angelo, R.~Di Salvo and A.~Fantini {\it et al.},
  Phys.\ Rev.\ Lett.\  {\bf 104} (2010) 241601
  [arXiv:1005.5230 [hep-ex]].

\bibitem{Gurzadyan:2010bt}
  V.~G.~Gurzadyan, V.~Bellini, M.~Beretta, J.~-P.~Bocquet, A.~D'Angelo, R.~Di Salvo, A.~Fantini and D.~Franco {\it et al.},
  arXiv:1004.2867 [physics.acc-ph].


\bibitem{Altschul:2006zz}
B.~Altschul,
  Phys.\ Rev.\ Lett.\  {\bf 98} (2007) 041603
  [hep-th/0609030].

  
\bibitem{Kaufhold:2007qd}
 C.~Kaufhold and F.~R.~Klinkhamer,
  Phys.\ Rev.\ D {\bf 76} (2007) 025024
  [arXiv:0704.3255 [hep-th]].



\bibitem{Klinkhamer:2007ak0}
 F.~R.~Klinkhamer and M.~Risse,
  Phys.\ Rev.\ D {\bf 77} (2008) 016002
  [arXiv:0709.2502 [hep-ph]].

\bibitem{Klinkhamer:2007ak1}  
F.~R.~Klinkhamer and M.~Risse,
  Phys.\ Rev.\ D {\bf 77} (2008) 117901
  [arXiv:0806.4351 [hep-ph]].

\bibitem{Eisele:2009zz}
 C.~.Eisele, A.~Y.~.Nevsky and S.~Schiller,
  Phys.\ Rev.\ Lett.\  {\bf 103} (2009) 090401.
 
\bibitem{Muller:2007zz}
 H.~Muller, P.~L.~Stanwix, M.~E.~Tobar, E.~Ivanov, P.~Wolf, S.~Herrmann, A.~Senger and E.~Kovalchuk {\it et al.},
  Phys.\ Rev.\ Lett.\  {\bf 99} (2007) 050401
  [arXiv:0706.2031 [physics.class-ph]].
 
 
\bibitem{Saathoff:2003zz}
 G.~Saathoff, S.~Karpuk, U.~Eisenbarth, G.~Huber, S.~Krohn, R.~M.~Horta, S.~Reinhardt and D.~Schwalm {\it et al.},
  Phys.\ Rev.\ Lett.\  {\bf 91} (2003) 190403.

\bibitem{NaturePhysics}
  S.~Reinhardt, G.~Saathoff, H.~Buhr, L.~A.~Carlson, A.~Wolf, D.~Schwalm, S.~Karpuk and C.~Novotny {\it et al.},
  Nature Phys.\  {\bf 3} (2007) 861.
  
 \bibitem{Tobar:2009gw}
   M.~E.~Tobar, E.~N.~Ivanov, P.~L.~Stanwix, J.~-M.~G.~le Floch and J.~G.~Hartnett,
  Phys.\ Rev.\ D {\bf 80} (2009) 125024
  [arXiv:0909.2076 [hep-ph]].
  

\bibitem{Hohensee:2010an}
    M.~A.~Hohensee, P.~.L.~Stanwix, M.~E.~Tobar, S.~R.~Parker, D.~F.~Phillips and R.~L.~Walsworth,
  Phys.\ Rev.\ D {\bf 82} (2010) 076001
  [arXiv:1006.1376 [hep-ph]].

\bibitem{SMENeutrino0}
  T.~Katori [MiniBooNE Collaboration],
  arXiv:1008.0906 [hep-ph].

\bibitem{SMENeutrino1}
 R.~Abbasi {\it et al.}  [IceCube Collaboration],
  Phys.\ Rev.\ D {\bf 82} (2010) 112003
  [arXiv:1010.4096 [astro-ph.HE]].

\bibitem{SMENeutrino2}
 P.~Adamson {\it et al.}  [MINOS Collaboration],
  Phys.\ Rev.\ Lett.\  {\bf 105} (2010) 151601
  [arXiv:1007.2791 [hep-ex]].

\bibitem{SMENeutrino3}  
  P.~Adamson {\it et al.}  [MINOS Collaboration],
  Phys.\ Rev.\ Lett.\  {\bf 101} (2008) 151601
  [arXiv:0806.4945 [hep-ex]].
  
\bibitem{SMENeutrino4} 
  L.~B.~Auerbach {\it et al.}  [LSND Collaboration],
  Phys.\ Rev.\ D {\bf 72} (2005) 076004
  [hep-ex/0506067].

\bibitem{SMENeutrino5}
  B.~Altschul,
  J.\ Phys.\ Conf.\ Ser.\  {\bf 173} (2009) 012003.

\bibitem{SMENeutrino6}
  V.~Barger, D.~Marfatia and K.~Whisnant,
  Phys.\ Lett.\ B {\bf 653} (2007) 267
  [arXiv:0706.1085 [hep-ph]].


  
 \bibitem{SMEMesons0}
   A.~Di Domenico {\it et al.}  [KLOE Collaboration],
  Found.\ Phys.\  {\bf 40} (2010) 852.

\bibitem{SMEMesons1}  
 A.~Di Domenico {\it et al.}  [KLOE Collaboration],
  J.\ Phys.\ Conf.\ Ser.\  {\bf 171} (2009) 012008.

 \bibitem{SMEMesons2}
 F.~Bossi {\it et al.}  [KLOE Collaboration],
  Riv.\ Nuovo Cim.\  {\bf 31} (2008) 531
  [arXiv:0811.1929 [hep-ex]].
  
 \bibitem{SMEMesons3}
 M.~Testa [KLOE Collaboration],
  arXiv:0805.1969 [hep-ex].
  
 \bibitem{SMEMesons4}
  D.~Babusci {\it et al.}  [KLOE-2 Collaboration],
  arXiv:1312.6818 [hep-ex].
  
 \bibitem{SMEMesons5}
 H.~Nguyen,
  hep-ex/0112046.

 \bibitem{SMEMesons6}
  V.~A.~Kostelecky,
  Phys.\ Rev.\ Lett.\  {\bf 80} (1998) 1818
  [hep-ph/9809572].
   
 \bibitem{SMEMesons7}
 J.~M.~Link, M.~Reyes, P.~M.~Yager, J.~C.~Anjos, I.~Bediaga, C.~Gobel, J.~Magnin and A.~Massafferri {\it et al.},
  Phys.\ Lett.\ B {\bf 556} (2003) 7
  [hep-ex/0208034].
   
 \bibitem{SMEMesons8}
  B.~Aubert {\it et al.}  [BaBar Collaboration],
  Phys.\ Rev.\ Lett.\  {\bf 100} (2008) 131802
  [arXiv:0711.2713 [hep-ex]].
   
 \bibitem{SMEMesons9}
  B.~Aubert {\it et al.}  [BaBar Collaboration],
  hep-ex/0607103.
   
 \bibitem{SMEMesons10}
B.~Altschul,
  Phys.\ Rev.\ D {\bf 77} (2008) 105018
  [arXiv:0712.1579 [hep-ph]].


\bibitem{SMEelectron0}  
B.~R.~Heckel, E.~G.~Adelberger, C.~E.~Cramer, T.~S.~Cook, S.~Schlamminger and U.~Schmidt,
  Phys.\ Rev.\ D {\bf 78} (2008) 092006
  [arXiv:0808.2673 [hep-ex]].
    
\bibitem{SMEelectron1}  
  B.~R.~Heckel, C.~E.~Cramer, T.~S.~Cook, E.~G.~Adelberger, S.~Schlamminger and U.~Schmidt,
  Phys.\ Rev.\ Lett.\  {\bf 97} (2006) 021603
  [hep-ph/0606218].

\bibitem{SMEelectron2}  
  L.~-S.~Hou, W.~-T.~Ni and Y.~-C.~M.~Li,
  Phys.\ Rev.\ Lett.\  {\bf 90} (2003) 201101
  [physics/0009012 [physics.space-ph]].

\bibitem{SMEelectron3}  
 H.~Dehmelt, R.~Mittleman, R.~S.~van Dyck, Jr. and P.~Schwinberg,
  Phys.\ Rev.\ Lett.\  {\bf 83} (1999) 4694
  [hep-ph/9906262].

\bibitem{SMEelectron4}  
R.~K.~Mittleman, I.~I.~Ioannou, H.~G.~Dehmelt and N.~Russell,
  Phys.\ Rev.\ Lett.\  {\bf 83} (1999) 2116.
    
 \bibitem{SMEelectron5}   
  V.~A.~Kostelecky and C.~D.~Lane,
  Phys.\ Rev.\ D {\bf 60} (1999) 116010
  [hep-ph/9908504].

\bibitem{SMEelectron6}  
  B.~Altschul,
  Phys.\ Rev.\ D {\bf 82} (2010) 016002
  [arXiv:1005.2994 [hep-ph]].
  
\bibitem{SMEelectron7}  
  B.~Altschul,
  Phys.\ Rev.\ D {\bf 81} (2010) 041701
  [arXiv:0912.0530 [hep-ph]].
  

\bibitem{SMEelectron8}  
 C.~D.~Lane,
  Phys.\ Rev.\ D {\bf 72} (2005) 016005
  [hep-ph/0505130].
  
\bibitem{SMEelectron9}  
B.~Altschul,
  Phys.\ Rev.\ D {\bf 74} (2006) 083003
  [hep-ph/0608332].
  
\bibitem{SMEelectron10}  
 B.~Altschul,
  Astropart.\ Phys.\  {\bf 28} (2007) 380
  [hep-ph/0610324].
  
\bibitem{SMEelectron11}  
F.~W.~Stecker and S.~L.~Glashow,
  Astropart.\ Phys.\  {\bf 16} (2001) 97
  [astro-ph/0102226].
  
\bibitem{SMEelectron12}  
 B.~Altschul,
  Phys.\ Rev.\ D {\bf 75} (2007) 041301
  [hep-ph/0612288].

\bibitem{SMEproton0}
J.~M.~Brown, S.~J.~Smullin, T.~W.~Kornack and M.~V.~Romalis,
  Phys.\ Rev.\ Lett.\  {\bf 105} (2010) 151604
  [arXiv:1006.5425 [physics.atom-ph]].

\bibitem{SMEproton1}
  M.~A.~Humphrey, D.~F.~Phillips, E.~M.~Mattison, R.~F.~C.~Vessot, R.~E.~Stoner and R.~L.~Walsworth,
  Phys.\ Rev.\ A {\bf 68} (2003) 063807
  [physics/0103068].
  
 \bibitem{SMEproton2} 
  D.~F.~Phillips, M.~A.~Humphrey, E.~M.~Mattison, R.~E.~Stoner, R.~F.~C.~Vessot and R.~L.~Walsworth,
  Phys.\ Rev.\ D {\bf 63} (2001) 111101
  [physics/0008230].

\bibitem{SMEproton3}
  P.~Wolf, F.~Chapelet, S.~Bize and A.~Clairon,
  Phys.\ Rev.\ Lett.\  {\bf 96} (2006) 060801
  [hep-ph/0601024].
  
\bibitem{SMEproton4}
 G.~Gabrielse, A.~Khabbaz, D.~S.~Hall, C.~Heimann, H.~Kalinowsky and W.~Jhe,
  Phys.\ Rev.\ Lett.\  {\bf 82} (1999) 3198.

\bibitem{SMEneutron0}
 C.~Gemmel, W.~Heil, S.~Karpuk, K.~Lenz, Y.~.Sobolev, K.~Tullney, M.~Burghoff and W.~Kilian {\it et al.},
  Phys.\ Rev.\ D {\bf 82} (2010) 111901
  [arXiv:1011.2143 [gr-qc]].

\bibitem{SMEneutron1}
  F.~Allmendinger, W.~Heil, S.~Karpuk, W.~Kilian, A.~Scharth, U.~Schmidt, A.~Schnabel and Y.~.Sobolev {\it et al.},
  arXiv:1312.3225 [gr-qc].
  
\bibitem{SMEneutron2}
  K.~Tullney, C.~Gemmel, W.~Heil, S.~Karpuk, K.~Lenz, Y.~.Sobolev, M.~Burghoff and W.~Kilian {\it et al.},
  arXiv:1008.0579 [hep-ph].

\bibitem{SMEneutron3}
I.~Altarev, C.~A.~Baker, G.~Ban, G.~Bison, K.~Bodek, M.~Daum, P.~Fierlinger and P.~Geltenbort {\it et al.},
  Phys.\ Rev.\ Lett.\  {\bf 103} (2009) 081602
  [arXiv:0905.3221 [nucl-ex]].
  
\bibitem{SMEneutron4}  
  V.~Flambaum, S.~Lambert and M.~Pospelov,
  Phys.\ Rev.\ D {\bf 80} (2009) 105021
  [arXiv:0902.3217 [hep-ph]].

\bibitem{SMEneutron5}
 B.~Altschul,
  Phys.\ Rev.\ D {\bf 79} (2009) 061702
  [arXiv:0901.1870 [hep-ph]].
  
\bibitem{SMEneutron6}
F.~Cane, D.~Bear, D.~F.~Phillips, M.~S.~Rosen, C.~L.~Smallwood, R.~E.~Stoner, R.~L.~Walsworth and V.~A.~Kostelecky,
  Phys.\ Rev.\ Lett.\  {\bf 93} (2004) 230801
  [physics/0309070].
  
\bibitem{SMEneutron7}
D.~Bear, R.~E.~Stoner, R.~L.~Walsworth, V.~A.~Kostelecky and C.~D.~Lane,
  Phys.\ Rev.\ Lett.\  {\bf 85} (2000) 5038
   [Erratum-ibid.\  {\bf 89} (2002) 209902]
  [physics/0007049].
  
\bibitem{SMEneutron8}
B.~Altschul,
  Phys.\ Rev.\ D {\bf 78} (2008) 085018
  [arXiv:0805.0781 [hep-ph]].

\bibitem{SMEneutron9}
 B.~Altschul,
  Phys.\ Rev.\ D {\bf 75} (2007) 023001
  [hep-ph/0608094].


\bibitem{Carone:2006tx}
  C.~D.~Carone, M.~Sher and M.~Vanderhaeghen,
  Phys.\ Rev.\ D {\bf 74} (2006) 077901
  [hep-ph/0609150].

\bibitem{Kostelecky:2008ts}
  V.~A.~Kostelecky and N.~Russell,
  Rev.\ Mod.\ Phys.\  {\bf 83} (2011) 11
  [arXiv:0801.0287 [hep-ph]].


\bibitem{Noscillations}
  J.~S.~Diaz and A.~Kostelecky,
  Phys.\ Lett.\  B {\bf 700} (2011) 25
  [arXiv:1012.5985 [hep-ph]].


\bibitem{Moscillations}
 A.~Kostelecky and R.~Van Kooten,
  Phys.\ Rev.\ D {\bf 82} (2010) 101702
  [arXiv:1007.5312 [hep-ph]].

\bibitem{Finsler}
   A.~Kostelecky,
  Phys.\ Lett.\ B {\bf 701} (2011) 137
  [arXiv:1104.5488 [hep-th]].

\bibitem{Tobar:2004vi}
  M.~E.~Tobar, P.~Wolf, A.~Fowler and J.~G.~Hartnett,
  Phys.\ Rev.\ D {\bf 71} (2005) 025004
   [Erratum-ibid.\ D {\bf 75} (2007) 049902]
  [hep-ph/0408006].



\bibitem{data10}
  J. H. Hough, C. Brindle,D. J. Axon, J. Bailey and W. B. Sparks,
  Mon.Not.Roy.Astron.Soc. {\bf 224} (1987) 1013.
  
\bibitem{data11}
  C. Brindle, J. H. Hough, J. A. Bailey, D. J.  Axon, M. J.  Ward, W. B. Sparks, I. S. McLean, 
  Mon.Not.Roy.Astron.Soc. {\bf 244} (1990) 577.

\bibitem{data12}  
  A. Cimatti, A.  Dey, W. van Breugel, R. Antonucci, H. Spinrad, 
  Astrophys.J. {\bf 465} (1996)145.

\bibitem{data13}
A. Dey , A.  Cimatti, W.  van Breugel, R. Antonucci, H. Spinrad,
Astrophys.J. {\bf 465} (1996) 157.

\bibitem{data14}
 A. Cimatti,  A. Dey, W. van Breugel, T. Hurt, R.  Antonucci,  
Astrophys.J. {\bf 476} (1997) 677.

\bibitem{data15}
M. S. Brotherton,  H. D. Tran, W. van Breugel, A.  Dey,  R. Antonucci,
Astrophys.J. {\bf 487} (1977) L113.%

\bibitem{data16}
M. S. Brotherton,
  B. J.  Wills,  A. Dey, W. van Breugel, R.  Antonucci, 
Astrophys.J. {\bf 501} (1998) 110. %

\bibitem{data17}
M. S. Brotherton, N.  Arav, R. H. Becker, H. D.  Tran, M. D.  Gregg, R. L. White, S. A. Laurent-Muehleisen, W. Hack,
Astrophys.J. {\bf 546} (2001) 134.

\bibitem{data18}
M. Kishimoto, R. Antonucci, A. Cimatti, T. Hurt, A. Dey, W. van Breugel, H. Spinrad, 
Astrophys.J. {\bf 547}(2001) 667.

\bibitem{data19}
J. Vernet, R. A. E. Fosbury, M. Villar-Mart?n, M. H. Cohen, A. Cimatti, S. di Serego Alighieri, R. W. Goodrich, 
Astron.Astrophys. {\bf 366} (2001) 7.


\bibitem{Mewes:2006ad}
M.~Mewes and A.~Petroff,
  Phys.\ Rev.\ D {\bf 75} (2007) 056002
  [hep-ph/0612372].

\bibitem{Michelson0}
  A.~A.~Michelson and E.~W.~Morley,
  Am.\ J.\ Sci.\  {\bf 34} (1887) 333.

\bibitem{Michelson1}
A.~A.~Michelson, and E.~W.~Morley, 
Philos. Mag. {\bf 24} (1887) 449-463.


\bibitem{Robertson}
 H.~P.~Robertson,
  Rev.\ Mod.\ Phys.\  {\bf 21} (1949) 378.
  
\bibitem{Mansouri}
 R.~Mansouri and R.~U.~Sexl,
  Gen.\ Rel.\ Grav.\  {\bf 8} (1977) 497.

\bibitem{Trimmer}
  W.~S.~N.~Trimmer, R.~F.~Baierlein, J.~E.~Faller and H.~A.~Hill,
  Phys.\ Rev.\ D {\bf 8} (1973) 3321
   [Erratum-ibid.\ D {\bf 9} (1974) 2489].
 
 \bibitem{Triangle}
 O.~D.~D.~Soares, 
 J. Phys. E: Sci. Instrum. {\bf 11} ( 1978) 773. 


\bibitem{Altschul:2009xh}
B.~Altschul,
  Phys.\ Rev.\ D {\bf 80} (2009) 091901
  [arXiv:0905.4346 [hep-ph]].

\bibitem{Kostelecky:2009zp}
V.~A.~Kostelecky and M.~Mewes,
  Phys.\ Rev.\ D {\bf 80} (2009) 015020
  [arXiv:0905.0031 [hep-ph]].



\bibitem{Bailey:2006fd}
Q.~G.~Bailey and V.~A.~Kostelecky,
  Phys.\ Rev.\ D {\bf 74} (2006) 045001
  [gr-qc/0603030].

\bibitem{Lambda}
  M.~Tegmark {\it et al.}  [SDSS Collaboration],
  Phys.\ Rev.\ D {\bf 69} (2004) 103501
  [astro-ph/0310723].

\bibitem{Baynes:2011nw}
F.~Baynes, A.~Luiten and M.~Tobar,
  Phys.\ Rev.\ D {\bf 84} (2011) 081101
  [arXiv:1108.5414 [gr-qc]].
  
\bibitem{Sagnac1}
 Georges Sagnac, 
 Comptes Rendus \textbf{157} (1913) 708-710.

\bibitem{Sagnac2} 
 Georges Sagnac,
 Comptes Rendus \textbf{157} (1913)1410-1413. 
 


\bibitem{GPS}
  P.~Wolf and G.~Petit,
  Phys.\ Rev.\ A {\bf 56} (1997) 4405.
  
\bibitem{Michimura:2013kca} 
  Y.~Michimura, N.~Matsumoto, N.~Ohmae, W.~Kokuyama, Y.~Aso, M.~Ando and K.~Tsubono,
  Phys.\ Rev.\ Lett.\  {\bf 110} (2013) 200401
  [arXiv:1303.6709 [gr-qc]].

\bibitem{Michimura:2013via}
  Y.~Michimura, M.~Mewes, N.~Matsumoto, Y.~Aso and M.~Ando,
  Phys.\ Rev.\ D {\bf 88} (2013) 111101
  [arXiv:1310.1952 [gr-qc]].
  
\end{thebibliography}
\end{document}